# Adapting sentiment analysis for tweets linking to scientific papers

Natalie Friedrich[1], Timothy D. Bowman[2], Wolfgang G. Stock[1] & Stefanie Haustein[2]

[1] *natalie.friedrich@hhu.de, stock@phil.hhu.de*
Heinrich Heine University Düsseldorf, Institute of Linguistics and Information, Department of Information Science, Düsseldorf (Germany)

[2] *stefanie.haustein@umontreal.ca, timothy.bowman@umontreal.ca*
École de bibliothéconomie et des sciences de l'information, Université de Montréal, Montréal (Canada)

**Introduction**

In the context of "altmetrics", tweets have been discussed as potential indicators of immediate and broader societal impact of scientific documents (Thelwall et al., 2013a). However, it is not yet clear to what extent Twitter captures actual research impact. A small case study (Thelwall et al., 2013b) suggests that tweets to journal articles neither comment on nor express any sentiments towards the publication, which suggests that tweets merely disseminate bibliographic information, often even automatically (Haustein et al., in press). This study analyses the sentiments of tweets for a large representative set of scientific papers by specifically adapting different methods to academic articles distributed on Twitter. Results will help to improve the understanding of Twitter's role in scholarly communication and the meaning of tweets as impact metrics.

**Dataset and Methods**

*Tweets and research articles*
The study is based on all articles and reviews published in 2012 in the Web of Science (WoS) linked to tweets via the Digital Object Identifier (DOI) as captured by Altmetric.com until 30 June 2014. The dataset consists of 663,547 original tweets (i.e., excluding retweets) mentioning 238,281 documents.

*Sentiment tools*
A sentiment represents an emotion expressed by a person based on their opinion towards a subject. Text-based sentiment analysis focuses largely on identifying positive and negative, as well as the absence of, sentiments using linguistic algorithms (Thelwall et al., 2010). For our purposes the sentiment expressed in a tweet linking to a scientific paper is assumed to reflect the opinion of the tweeting user towards the paper. SentiStrength[1] ($s_1$) and Sentiment140[2] ($s_2$) were selected to automatically detect sentiments. SentiStrength assigns values from -5 to +5 to certain terms in a lexicon. Each processed tweet receives a negative and a positive value. To assign each tweet to exactly one category (positive, negative, neutral), the stronger value determines the sentiment. Sentiment140 provides one sentiment value per tweet on a scale from 0 (negative) to 4 (positive). For better comparison values are converted to obtain three sentiment categories positive, negative, and neutral. While SentiStrength has been developed for short online texts and Sentiment140 was particular implemented to analyse tweets, none of the tools seem suited to analyse tweets related to scientific topics. In contrast to SentiStrength, which provides options to change the lexicon, Sentiment140 is less transparent and only allows insight into the training corpus.

*Intellectual coding of sentiments*
The text from 1,000 random tweets was analysed and compared to the title of the papers the tweets linked to in order to gain an understanding of the discussions of scientific papers on Twitter and to determine their sentiment intellectually $s_i$. A second intellectual assessment is undertaken with regard to the capabilities of the sentiment analysis tools. For example, Natural Language Processing (NLP) tools are not able to detect irony. The results of these assessments function as the ground truth $s_0$, to which sentiments detected by the tools are compared.

*Cleaning tweets*
A tweet consists of 140 characters including text, hashtags (following the # sign), user names (following the @ sign), and/or links to websites. As user names, URLs, and the # sign are not considered to be part of the tweet content regarding the sentiment analysis, they were removed from the tweet. Hashtag terms are kept as they are assumed to carry meaning and sentiment. The tweets without specific affordances are called $t_0$.

The intellectual analysis revealed that many tweets contained the title of the scientific paper to which they linked, which influences the sentiment analysis—even though it does not reflect the users emotion and opinion towards the paper. As the sentiment tools are not adapted to scientific language, certain research topics are assigned positive or negative sentiments. For example, in SentiStrength the term 'cancer' receives the value -4 and 'disease' -3. As this influences the outcome of the sentiment analysis, tweets $t_0$ were further adapted by removing all title terms from the particular paper to which they link (using regular expressions in PHP) to derive tweets adapted for sentiment analysis $t_a$.

---
[1] http://sentistrength.wlv.ac.uk/
[2] http://help.sentiment140.com/home

In addition to removing title words from tweets to avoid false positives regarding the sentiment detection, the lexicon was adapted to the scientific context for SentiStrength by identifying the terms leading to disagreement between $s_0$ and $s_1$. Overall, 51 terms (e.g., 'cancer', 'disease' or 'obesity' for negative sentiments, 'baby' or 'care' for positive sentiments) were removed from the lexicon. Results for SentiStrength after the lexicon changes are denoted as $s'_1$. The lexicon for Sentiment140 was not accessible and thus could not be adapted.

Results obtained by SentiStrength ($s_1$ and $s'_1$) and Sentiment140 $s_2$ are compared to the ground truth $s_0$ for cleaned tweets $t_0$ and $t_a$ using percentage overlap and Cohen's Kappa κ.

**Preliminary Results**

The intellectual assessment of the tweet content $s_i$ identified 4.3% of the 1,000 random tweets to contain positive, 0.9% negative, and 94.8% neutral sentiment, which is in agreement with findings by Thelwall et al. (2013b).

**Table 1. Intellectual ($s_0$) and automated ($s_1$, $s'_1$, $s_2$) sentiment detection for 1,000 tweets.**

|       |        | Sentiments (%) |      |      | Agreement w/ $s_0$ |      |
|-------|--------|----|------|------|------|------|
|       |        | +  | –    | n    | %    | κ    |
|       | $s_i$  | 4.3 | 0.9 | 94.8 | n/a  | n/a  |
|       | $s_0$  | 4.1 | 0.6 | 95.3 | n/a  | n/a  |
| $t_0$ | $s_1$  | 12.2 | 33.8 | 54.0 | 56.8 | 0.10 |
|       | $s_2$  | 0.6 | 1.6 | 97.8 | 94.3 | 0.16 |
| $t_a$ | $s_1$  | 8.2 | 11.2 | 80.6 | 83.8 | 0.29 |
|       | $s'_1$ | 8.0 | 2.8 | 89.2 | 92.9 | 0.52 |
|       | $s_2$  | 0.7 | 1.0 | 98.3 | 94.6 | 0.14 |

Results for SentiStrength ($s_1$, $s'_1$) and Sentiment140 ($s_2$) compared to the ground truth $s_0$ are shown in Table 1. Removing paper title terms from the tweets increases the accuracy in particular for neutral and positive tweets and raises agreement with $s_0$ from 56.8% to 83.8% for $s_1$, representing fair agreement according to Cohen's Kappa (κ=0.29). The process of adapting the lexicon ($s'_1$) leads to an additional increase to 92.9% (κ =0.52, moderate agreement). 90.2% of 41 positive tweets and 93.2% of 953 neutral tweets are detected correctly by $s'_1$ for $t_a$. However, the detection of negative sentiments decreases from 100% ($s_1$) to 66.7% ($s'_1$), as only 4 of 6 negative tweets were identified by $s'_1$.

Although the overall agreement between $s_2$ and $s_0$ for $t_0$ represents 94.3%, only 14.6% positive sentiments and none of the 6 negative sentiments were detected correctly by Sentiment140. The high overall agreement arises from the agreement of neutral sentiment that yields 937 tweets. Removing the title words from tweets leads to a small increase of the overall percentage agreement for Sentiment140 to 94.6%, however the percentage of identified positive tweets decreases to 12.2%.

**Discussion and Future Work**

Our analysis shows that current sentiment tools are not able to accurately detect sentiments for the specific context of tweets discussing academic papers. While SentiStrength overestimates sentiments of tweets about scientific papers, Sentiment140 is not able to detect any negative tweets and only 14.6% of positive tweets leading to slight agreement (κ=0.16). As it does not allow access to the lexicon, Sentiment140 remains a black box.

Automatic sentiment detection was significantly improved for SentiStrength by adjusting tweets (removing title terms) and lexicon leading from slight (κ=0.10) to moderate agreement (κ=0.52). However, the detection of negative sentiments remains problematic.

Future work will focus on improving negative sentiment detection by analyzing specific cases of false positives. The aim is to develop an adapted lexicon in order to perform an sentiment analysis the 663,547 tweets linking to 238,281 documents.

**References**

Haustein, S., Bowman, T. D., Holmberg, K., Tsou, A., Sugimoto, C. R., & Larivière, V. (in press). Tweets as impact indicators: Examining the implications of automated bot accounts on Twitter. *Journal of the Association for Information Science and Technology*. Retrieved from http://arxiv.org/abs/1410.4139

Thelwall, M., Haustein, S., Lariviére, V., Sugimoto, C.R. (2013a). Do Altmetrics work? Twitter and Ten Other Social Web Services. *PLoS ONE 8(5)*: e64841.

Thelwall, M., Tsou, A., Weingart, S., Holmberg, K., & Haustein, S. (2013b). Tweeting links to academic articles. *Cybermetrics: International Journal of Scientometrics, Informetrics and Bibliometrics*, 1–8. Retrieved from http://cybermetrics.cindoc.csic.es/articles/v17i1p1.html

Thelwall, M., Buckley, K., Paltoglou, G. Cai, Kappas, A. (2010). Sentiment strength detection in short informal text. *Journal oft he American Society for Information Science und Technology,* 61(12), 2544-2558.